\documentclass[twocolumn]{aastex631}
\usepackage[utf8]{inputenc}
\usepackage{natbib}
\usepackage{graphicx}
\usepackage{xcolor}
\usepackage{threeparttable}
\usepackage{multirow}
\usepackage{amsmath}

\received{}
\revised{}
\accepted{}
\submitjournal{AAS Journals}

\shorttitle{On the Ariel Tier System}
\shortauthors{Radica et al.}

\defcitealias{Mugnai_Pascale_Edwards_Papageorgiou_Sarkar_2020}{M20}

\begin{document}

\title{On the Information Content of Ariel Transmission Spectra: Reassessing the Tier System}

\correspondingauthor{Michael Radica}
\email{radicamc@uchicago.edu}

\author[0000-0002-3328-1203]{Michael Radica}
\altaffiliation{NSERC Postdoctoral Fellow}
\affiliation{Department of Astronomy \& Astrophysics, University of Chicago, 5640 South Ellis Avenue, Chicago, IL 60637, USA}
\affiliation{Institut Trottier de Recherche sur les Exoplanètes, 1375 Avenue Thérèse-Lavoie-Roux, Montréal, QC H2V 0B3, Canada}

\author[0000-0001-6129-5699]{Nicolas B.\ Cowan}
\affiliation{Department of Physics, McGill University, 3600 rue University, Montréal, QC, H3A 2T8, Canada}
\affiliation{Department of Earth \& Planetary Sciences, McGill University, 3450 rue University, Montréal, QC H3A 0E8, Canada}
\affiliation{Institut Trottier de Recherche sur les Exoplanètes, 1375 Avenue Thérèse-Lavoie-Roux, Montréal, QC H2V 0B3, Canada}

\author[0000-0001-5383-9393]{Ryan Cloutier}
\affiliation{Department of Physics \& Astronomy, McMaster University, 1280 Main Street W, Hamilton, ON L8S 4L8, Canada}

\author[0009-0009-3021-6156]{Leo Yang Wang}
\affiliation{Department of Astronomy, Cornell University, 122 Sciences Drive, Ithaca, NY 14853, USA}
\affiliation{Department of Physics \& Astronomy, McMaster University, 1280 Main Street W, Hamilton, ON L8S 4L8, Canada}

\begin{abstract}
The European Space Agency's Ariel mission will conduct a survey of the atmospheric properties of exoplanets around bright stars. The mission is nominally divided into three Tiers. The Tier 1 survey will consist of low-precision observations of $\sim$1000 planets, with a subset of these included in the higher-precision Tier 2 survey expected to be necessary for atmospheric characterization. Tier 3 will be repeated observations of a small number of benchmark planets. Though previous studies have assessed the ability of Ariel to uncover population-level trends, they have generally presupposed a given Tier. Here we interrogate this assumption and assess the information content of Ariel transmission spectra as a function of Tier for three benchmark planets: a hot-Saturn, warm-Neptune, and temperate sub-Neptune. We simulate a grid of Ariel transit spectra at different Tiers for each target and use retrievals to assess which chemical species are well constrained. We find that for giant planets like a hot-Saturn or warm-Neptune, Tier 1-quality observations are sufficient for $\lesssim$1.5\,dex constraints on H$_2$O and CO$_2$, even in the presence of grey clouds --- meaning important chemical insights are already obtainable in the Tier 1 survey. Moving to Tiers 2 and 3 results in an incremental increase in precision as well as other molecules becoming constrainable in certain scenarios (e.g., H$_2$S, CO). Tier 1 observations are also sufficient to identify CH$_4$ in a temperate sub-Neptune, whereas observations with at least Tier 2 precision are necessary for CO$_2$. The number of transits necessary to reach this observational S/N, however, may be prohibitive for the inclusion of temperate sub-Neptunes in even the Tier 1 survey. 
\end{abstract}


\section{Introduction} 
\label{sec: Introduction}

The exoplanet population offers the opportunity to study the physical and chemical processes that govern planet formation and evolution on a much larger scale than is possible with the Solar System alone. Scheduled to launch in 2031, the European Space Agency's (ESA) Ariel mission is designed specifically to capitalize on this opportunity by performing a uniform survey of exoplanets and their atmospheres to uncover population-level trends \citep{tinetti_chemical_2018, zellem_constraining_2019, tinetti_ariel_2022, d'aoust_testing_2025}. 

It has long been recognized that population-level studies hold the key to uncovering many of the key processes that shape the exoplanet population. For example, the present-day atmospheric composition of exoplanets contain clues into the manner in which they form and evolve \citep[e.g.,][]{oberg_effects_2011, madhusudhan_towards_2014, booth_chemical_2017, penzlin_bowie_2024}. Interpreting these clues, though, has proven to be challenging as multiple different formation and evolutionary scenarios can often map to the same present day composition \citep[e.g.,][]{chachan_breaking_2023, penzlin_bowie_2024, feinstein_linking_2025}. However, population-level studies provide the opportunity to cut through the degeneracies of individual targets and uncover the processes that affect the aggregate sample \citep[e.g.,][]{mordasini_imprint_2016, welbanks_massmetallicity_2019, penzlin_bowie_2024, lothringer_library_2026}. Predicted to observe the atmospheres of $\sim$1000 exoplanets, it is in this area that Ariel will shine. 

\begin{figure*}
    \centering
    \includegraphics[width=0.95\linewidth]{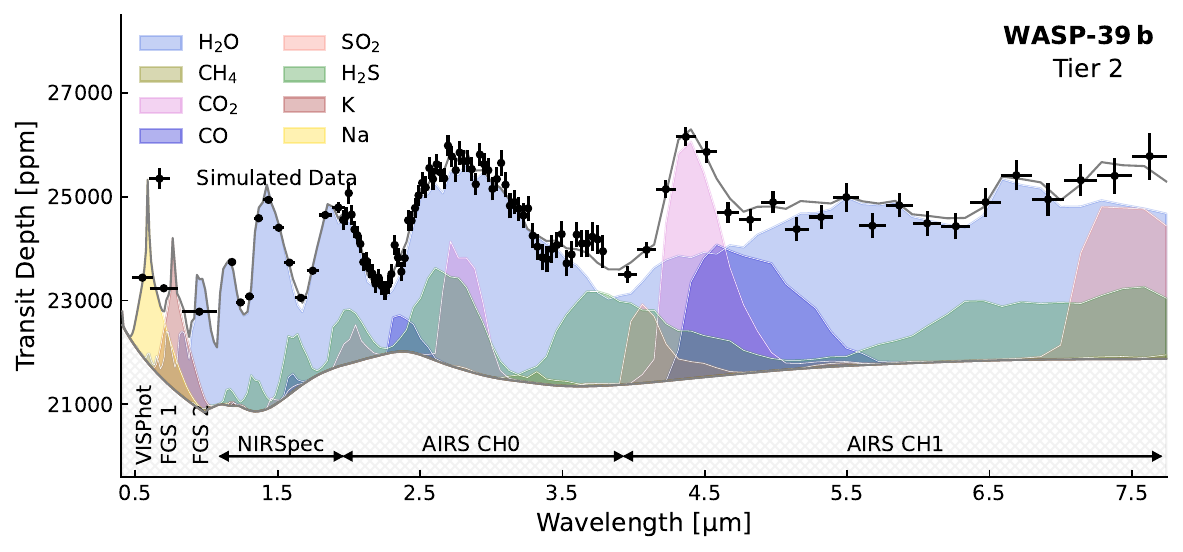}
    \caption{Example full-resolution Ariel transmission spectrum of the hot-Saturn WASP-39\,b simulated at Tier 2 S/N (black error bars). The underlying atmosphere model assumes chemical equilibrium with a solar C/O ratio (0.54), 10$\times$ solar metallicity, and no clouds. The opacity contributions of individual species to the total spectrum are shown with different colours and represent the major chemical species with significant opacity in the Ariel wavelength range. The wavebands of Ariel's instruments are labelled underneath the spectrum.}
    \label{fig: Spectrum Contributions}
\end{figure*}

One of the key advantages of Ariel is its ability to obtain atmospheric spectra with wide, simultaneous wavelength coverage. Ariel will have six science instruments, three photometers: VISPhot (0.5 -- 0.6\,µm), FGS1 (0.6 -- 0.81\,µm), and FGS2 (0.19 -- 1.1\,µm); as well as three low resolution spectrographs: NIRSpec (1.1 -- 1.95\,µm at $R$=20), AIRS Ch0 (1.93 -- 3.9\,µm at $R$=100), and AIRS Ch1 (3.9 -- 7.8\,µm, at $R$=30) (Figure~\ref{fig: Spectrum Contributions}; \citealp{edwards_updated_2019}). These six instruments will observe simultaneously, resulting in instantaneous 0.5 -- 7.8\,µm wavelength coverage --- wider than any single JWST instrument, albeit at lower S/N and spectral resolution \citep{changeat_synergetic_2025}. Figure~\ref{fig: Spectrum Contributions} shows an example simulated Ariel transmission spectrum of a hot-Saturn exoplanet, highlighting the plethora of chemical species which have significant opacity in the Ariel bandpass. 

Another important characteristic of Ariel's planned atmospheric surveys are their uniformity. Building on the work of \citet{zingales_ariel_2018}, \citet{edwards_updated_2019} and \citet{edwards_ariel_2022} constructed a list of possible targets for Ariel, vetting $\sim$2000 planets from hot-Jupiters to super-Earths. It is from this target list that planets will be selected for the Ariel Mission Reference Sample (MRS) to actually be observed \citep{tinetti_chemical_2018}. The exact number of planets to include in the MRS, as well as the ideal method for their selection, has yet to be finalized \citep[e.g.,][]{edwards_ariel_2022, cowan_maximizing_2025, panek_balancing_2026}.

\subsection{The Ariel Tier System}
\label{sec: The Tier System}

Once the MRS is set, Ariel's surveys will be conducted with a ``tiered'' approach \citep{tinetti_ariel_2022, edwards_updated_2019}. Following \citet{edwards_updated_2019}, Tier 1 will consist of low-signal-to-noise (S/N) observations to nominally constrain orbital parameters and first-order atmospheric properties (e.g., presence of molecular features, degree of cloudiness). From the Tier 1 sample, a substantial subset will be re-observed in Tier 2, which will consist of a series of follow-up observations to enable precision spectroscopic characterization. Finally, Tier 3 will enable repeated observations of a small selection of benchmark targets for in-depth characterization. There will also potentially be a fourth tier for bespoke observing strategies which do not fit into the transit survey, e.g., phase curves \citep{charnay_survey_2022} or eclipse mapping \citep{valentine_eclipse_2025}.

The three tiers of the transit survey are determined by achievable spectroscopic S/N \citep{tinetti_chemical_2018, edwards_updated_2019}. That is, a spectrum is considered to be at Tier X (where X $\in$ [1, 3]) S/N when it achieves a S/N $\geq$ 7 on atmospheric features at the spectral binning prescribed for Tier X. Lower tiers require coarser bins, i.e., $R\sim$~1, 3, and 1 for NIRSpec, AIRS Ch0, and Ch1, respectively at Tier 1; 10, 50, and 10 at Tier 2; and native resolutions at Tier 3. For further information on the specific Ariel Tier definitions, see \citet{tinetti_chemical_2018} and \citet{edwards_updated_2019}. Critically, all spectra observed at a given tier will have comparable S/N, whereas the number of transits necessary to reach said S/N will vary between targets.  

We reemphasize here, as this is a common source of confusion, that the Ariel Tier refers fundamentally to the \textit{S/N} of the observations and not the \textit{binning}. Throughout this work, whenever we quote the Tier of an observation we are referring to its S/N rather than binning. 

Several previous works have looked to assess the ability of Ariel to uncover underlying population-level trends with its surveys, however, these studies have generally assumed a specific Tier a priori for their simulated observations. For example, \citet{changeat_alfnoor_2020} assess detection limits for various molecules and the ability of an Ariel survey to uncover trends in, e.g., the abundance of H$_2$O with temperature. But they explicitly assume that Tier 2 or Tier 3 quality observations are necessary for this endeavour and do not explore Tier 1. Similarly, \citet{mugnai_alfnoor_2021} only analyze Tier 1 spectra for their ability to distinguish between clear and featureless spectra. \citet{barstow_retrieval_2022} compare the performance of various retrieval codes on simulated Ariel transit spectra, but again only consider Tier 2 quality simulations. \citet{zellem_constraining_2019} consider Ariel's ability to constrain mass-metallicity trends across a wider range of precisions, but do not use the standard tiered approach and instead assume a fixed number of transits. 

Although population-level analyses are ultimately the goal of the Ariel mission, analyzing trends presupposes the ability to constrain atmospheric species in individual planets. Here, we seek to rectify this gap in the existing Ariel literature by assessing the detectability of various molecules in exoplanet atmospheres with Ariel as a function of observational S/N (i.e., Ariel Tier). In this exploratory study, we limit ourselves to a small selection of benchmark planets which span the range of planetary and stellar types that make up the Ariel Target List. 

The structure of this work is as follows. In Section~\ref{sec: Simulations} we describe our atmosphere and instrument simulations. We then outline our retrieval analyses in Section~\ref{sec: Retrievals}, and discuss the results in Section~\ref{sec: Results}. Finally, we conclude in Section~\ref{sec: Conclusions}.

\section{Simulated Observations} 
\label{sec: Simulations}

\subsection{Atmosphere Forward Models}
\label{sec: Forward Models}

The goal of this study is to assess how well the abundances of various chemical species in exoplanetary atmospheres can be constrained with Ariel as a function of the achieved Tier. For this exploratory study we limit ourselves to benchmark planets spanning the range of systems potentially observable with Ariel. We choose three targets: WASP-39\,b, a hot-Saturn orbiting a G-type star; HAT-P-11\,b, a warm-Neptune orbiting a K-type star; and K2-18\,b, a temperate sub-Neptune orbiting an early M dwarf. The physical parameters of each target are summarized in Table~\ref{tab: Planet Parameters}.

\begin{deluxetable}{l|ccc}
 \centering
 \tabletypesize{\small}
 \label{tab: Planet Parameters}
 \tablecaption{Physical Parameters of Planetary Systems Relevant for This Work}
 \tablehead{Parameter & WASP-39\,b & HAT-P-11\,b & K2-18\,b}
    \startdata
     $J$ [Mag] & 10.663 & 7.608 & 9.763 \\
     R$_*$ [R$_\odot$] & 0.939 $^1$ & 0.770 $^3$ & 0.411 $^4$ \\
     St.\ Type & G8V $^2$ & K4V $^3$ & M2.5V $^4$ \\
     St.\ T$\rm _{eff}$ [K] & 5485 $^1$ & 4708 $^3$ & 3457 $^5$ \\
     St.\ [Fe/H] & 0.01 $^1$ & 0.29 $^3$ & 0.123 $^7$\\
     St.\ log\,g [cm/s$^2$] & 4.453 $^1$ & 4.37 $^3$ & 4.858 $^8$\\
     \rule{0pt}{3ex}R$\rm _p$ [R$\rm _{J}$] & 1.279 $^1$ & 0.43 $^3$ & 0.212 $^5$ \\
     M$\rm _p$ [M$\rm _{J}$] & 0.281 $^1$ & 0.09 $^3$ & 0.025 $^6$\\
     T$\rm _{eq}$ [K] & 1166 $^1$ & 878 $^3$ & 235 $^5$ \\
    \enddata
    \tablecomments{ $^1$ \citet{mancini_gaps_2018};
    $^2$ \citet{faedi_wasp-39b_2011};
    $^3$ \citet{stassun_accurate_2017};
    $^4$ \citet{benneke_spitzer_2017}
    $^5$ \citet{cloutier_characterization_2017}
    $^6$ \citet{radica_revisiting_2022}
    $^7$ \citet{benneke_spitzer_2017}
    $^8$ \citet{crossfield_candidates_2016}.
    }
\end{deluxetable}

We use the open source code \texttt{POSEIDON} \citep{macdonald_hd_2017, macdonald_poseidon_2023} to create atmosphere forward models for each of the three planets. We generate a plane-parallel atmosphere model with 100 layers spanning 2 to $-$7\,bar in log pressure at a resolution of $R$=10\,000. We assume a reference pressure of 10\,bar at which the planet radius is set. The temperature structure is assumed to be isothermal at the planet's equilibrium temperature. We set the chemical composition of each atmosphere as follows: for WASP-39\,b and HAT-P-11\,b we assume that the atmosphere is in chemical equilibrium at 10$\times$ and 50$\times$ solar metallicity respectively. These values are roughly consistent with previous atmosphere studies of each planet as well as broader mass-metallicity trends in transiting exoplanet atmospheres \citep[e.g.,][]{welbanks_massmetallicity_2019, constantinou_early_2023, feinstein_early_2023}. We assume a solar C/O ratio \citep[0.54;][]{asplund_chemical_2009} for WASP-39\,b \citep{alderson_early_2023, constantinou_early_2023, lueber_information_2024} and C/O=0.7 for HAT-P-11\,b to allow for more diversity of C-bearing species as might be possible for Neptune-sized planets \cite[e.g.,][]{moses_compositional_2013, radica_muted_2024, ashtari_heat_2026}. 

We consider opacity from the following chemical species for WASP-39\,b: H$_2$O \citep{polyansky_exomol_2018}, CO$_2$ \citep{yurckenko_exomol_2020}, CO \citep{li_rovibrational_2015}, CH$_4$ \citep{yurchenko_exomol_2024}, SO$_2$ \citep{underwood_exomol_2016}, H$_2$S \citep{azzam_exomol_2016}, NH$_3$ \citep{coles_exomol_2019}, Na \citep{ryabchikova_major_2015}, K \citep{ryabchikova_major_2015}, as well as collisionally-induced absorption (CIA) from H$_2$-H$_2$ \citep{Chubb2021} and H$_2$-He \citep{Chubb2021}. We use the same opacities, with the exception of SO$_2$, for HAT-P-11\,b. 

For each species, we use \texttt{FastChem} \citep{stock_fastchem_2018} to generate a volume mixing ratio (VMR) profile under chemical equilibrium and average the VMR over 10$^{-2}$--10$^{-5}$\,bar in pressure, which roughly corresponds to the pressures probed by transit observations. We then generate an atmosphere forward model assuming a vertically constant abundance profile at these values. This allows for easier comparison between input and retrieved values for individual chemical species, while still ensuring that the injected abundances remain physically grounded. For WASP-39\,b, we also increase the SO$_2$ VMR to 10$^{-6}$ to better match the the planet's real photochemically-enhanced SO$_2$ abundance \citep{tsai_photochemically_2023, powell_sulfur_2024}.

For K2-18\,b we use a bespoke composition based on the findings of \citet{madhusudhan_carbon-bearing_2023}. We include a simpler set of molecules: H$_2$O, CO$_2$, CH$_4$, CO, NH$_3$, and HCN \citep{barber_exomol_2014}, though the overall atmosphere setup remains the same as above. We inject vertically constant abundances of CH$_4$ and CO$_2$ based on the ``two-offsets'' retrieval in \citet{madhusudhan_carbon-bearing_2023}, and values consistent with their 3-$\sigma$ upper limits for all other molecules. For brevity, the injected VMRs for each planet are summarized in Table~\ref{tab: Injected Abundances}.

Finally, for each planet we simulate both a cloudy and cloud-free scenario. For the cloudy case we place an opaque and spatially uniform, grey cloud deck at 10$^{-3}$\,bar, which is sufficiently high in the atmosphere to significantly mute spectral features, but not erase them entirely (e.g., Figure~\ref{fig: Planet Spectra}). We also include a modified Rayleigh scattering slope, with an enhancement factor $\alpha$ and scattering slope $\gamma$, where $\gamma=-4$ is H$_2$ Rayleigh scattering \citep[e.g.,][]{macdonald_hd_2017, pinhas_retrieval_2018}. $\gamma = -4$ and $\alpha = 1.0$ in the forward model. We emphasize that these compositions are simply illustrative of the possible compositions of targets that could be studied with Ariel, and not perfect reproductions of any of the three planets.

\subsection{Ariel Instrument Simulations}
\label{sec: Instrument Simulations}

We use an ad-hoc noise model based on the ArielRad simulator 
\citep[][hereafter \citetalias{Mugnai_Pascale_Edwards_Papageorgiou_Sarkar_2020}]{Mugnai_Pascale_Edwards_Papageorgiou_Sarkar_2020} to generate synthetic transmission spectra. Our model scales the chromatic noise properties of a reference star presented in ArielRad \citepalias[i.e. GJ 1214;][]{Mugnai_Pascale_Edwards_Papageorgiou_Sarkar_2020} to those of a user-defined star.
GJ 1214 is a characteristic faint target for Ariel \citepalias{Mugnai_Pascale_Edwards_Papageorgiou_Sarkar_2020}, and therefore, the noise budget for GJ 1214 is dominated by photon noise rather than by other instrumental noise sources (e.g., dark current, readout, gain, and zodiacal background noise). For the systems considered in this study, our noise calculations include photon noise as well as achromatic noise sources from the payload noise floor and gain noise. For a one hour integration, we assume a signal noise floor of $p_0=20\,\mathrm{ppm}$ \citep{greene_characterizing_2016} and $p_g=40\,\mathrm{ppm}$ of the signal for the gain noise \citep{Baraffe_Homeier_Allard_Chabrier_2015}. As these noise sources do not vary with wavelength, their quadrature sum comprises the fixed noise term in our model; $p=\sqrt{p_0^2+p_g^2}$. 

To calculate the photon noise contributions, we begin by modelling the spectrum of each star using the PHOENIX stellar atmosphere grid \citep{husser_new_2013}. We select the model spectrum with the closest effective temperature, surface gravity, and metallicity values from Table~\ref{tab: Planet Parameters}. 
We linearly interpolate each model to a custom wavelength grid, $\lambda'$, set by the spectral resolving power of each Ariel instrument. We then rescale the flux density values by

\begin{equation}
S(\lambda')= S_{\mathrm{raw}}(\lambda') \left( R_\star/d \right)^2,
\end{equation}

\noindent where $R_\star$ is the stellar radius, $d$ is the distance to the system, and $S_{\mathrm{raw}}(\lambda)$ is the raw PHOENIX model spectrum in units of erg/s/cm$^2$/cm. While the resulting S/N of our synthetic observations needs to be calculated in units of photoelectrons, we ignore the conversion from flux density to photoelectron rate because the necessary scaling parameters (e.g., $E=hc/\lambda',\, \Delta \lambda',$ quantum efficiency, etc.) are shared between the reference and target stars, and therefore cancel out.




We proceed with calculating the photon noise budget for our target stars by scaling the noise for an ArielRad simulated observation of GJ 1214 \citepalias{Mugnai_Pascale_Edwards_Papageorgiou_Sarkar_2020}. The reference noise values from ArielRad, $\sigma_{\rm ref}(\lambda')$, are calculated for a fixed, one hour integration and are reported relative to the stellar signal $S_{\rm GJ1214}(\lambda')$ in units of $\sqrt{\mathrm{hr}}$ \citepalias{Mugnai_Pascale_Edwards_Papageorgiou_Sarkar_2020}. We use the PHOENIX model grid and the stellar parameters for GJ 1214 from \citet{Mahajan_Eastman_Kirk_2024} to model $S_{\rm GJ1214}(\lambda')$. We compute the resulting errors in our simulated transit light curves $\sigma(\lambda')$ by rescaling the reference noise model by $T_{14}/{\rm 1 hr}$; the in-transit integration time in hours, which we approximate as the planet's transit duration. 
Our resulting noise model is

\begin{equation}
    \sigma^2(\lambda') = \left( \frac{S(\lambda')}{S_{\rm GJ1214}(\lambda')} \right) \left( \frac{T_{14}}{\rm 1hr} \right)\, (\sigma_{\rm GJ1214}^2(\lambda') - p^2) + p^2,
    \label{eq:sigma}
\end{equation}

\noindent which yields the relative error in each light curve measurement, where the error is assumed to be fixed for each measurement. We proceed by converting this to a transit depth measurement following the Fisher information analysis of a piecewise linear transit model. Following \citet{Carter_Yee_Eastman_Gaudi_Winn_2008}, we define the variables

    \label{eq: transit depth variables}
\begin{align}
    \theta &= \frac{T_{12}}{T_{14}}, \\
    Q(\lambda') &= \sqrt{\Gamma T_{14}}\, \left( \frac{\delta}{\sigma(\lambda')} \right),
\end{align}

\noindent where $T_{12}$ is the ingress (or egress) duration, $T_{14}$ is the full transit duration, $\Gamma=1/{\rm 60}$\,sec is the assumed sampling rate, $\delta$ is the fractional transit depth, and $\sigma(\lambda')$ follows from Eq.~\ref{eq:sigma}.  We assume the limit of small impact parameter, which simplifies $\theta$ to $\theta\approx \sqrt{\delta}$, and then calculate the error in transit depth using the relation

\begin{equation}
    \label{eq: depth error equation}
    \sigma_\delta(\lambda')=\sqrt{\frac{1}{Q(\lambda')^2} \frac{\delta^2}{1-\theta}}.
\end{equation}

\noindent Finally, we inflate the errors by 10\% to allow for the fact that, in practice, observations may not always reach the predicted photon noise level. Uncertainties roughly 10\% above photon noise is a value typically found for JWST transmission spectra \citep[e.g.,][]{radica_awesome_2023, alderson_jwst_2024, radica_super-solar_2026}.

\begin{figure}
    \centering
    \includegraphics[width=0.95\linewidth]{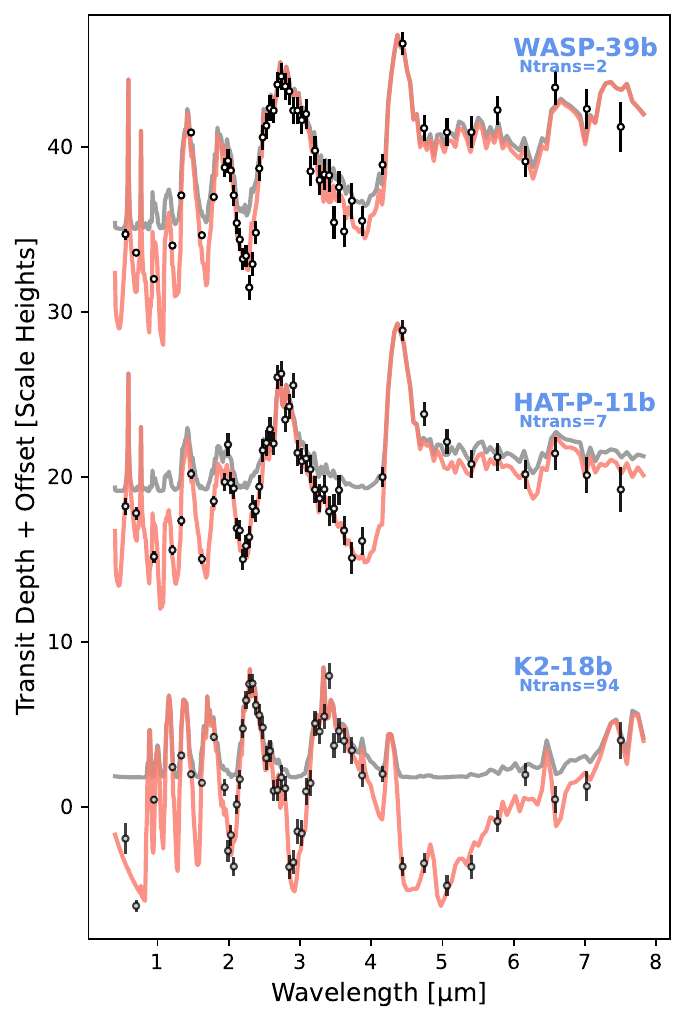}
    \caption{Simulated cloud-free atmosphere spectra and Tier 2 Ariel observations for the three planets considered in this study. The number of transits required to reach Tier 2 S/N is noted for each planet. Simulated observations are shown with black error bars (wavelength bin widths are omitted for visual clarity), and are binned to $R$=10, $R$=50, and $R$=10 (i.e., Tier 2 binning) for NIRSpec, AIRS Ch0, and AIRS Ch1, respectively. Binning is for plotting purposes only and the retrieval analyses are performed on the native resolution observations. The underlying atmosphere spectra are overplotted in red. Spectra are given an arbitrary vertical offset for ease of visualization. For comparison purposes, cloudy models are also shown in grey.}
    \label{fig: Planet Spectra}
\end{figure}

\begin{figure}
    \centering
    \includegraphics[width=0.95\linewidth]{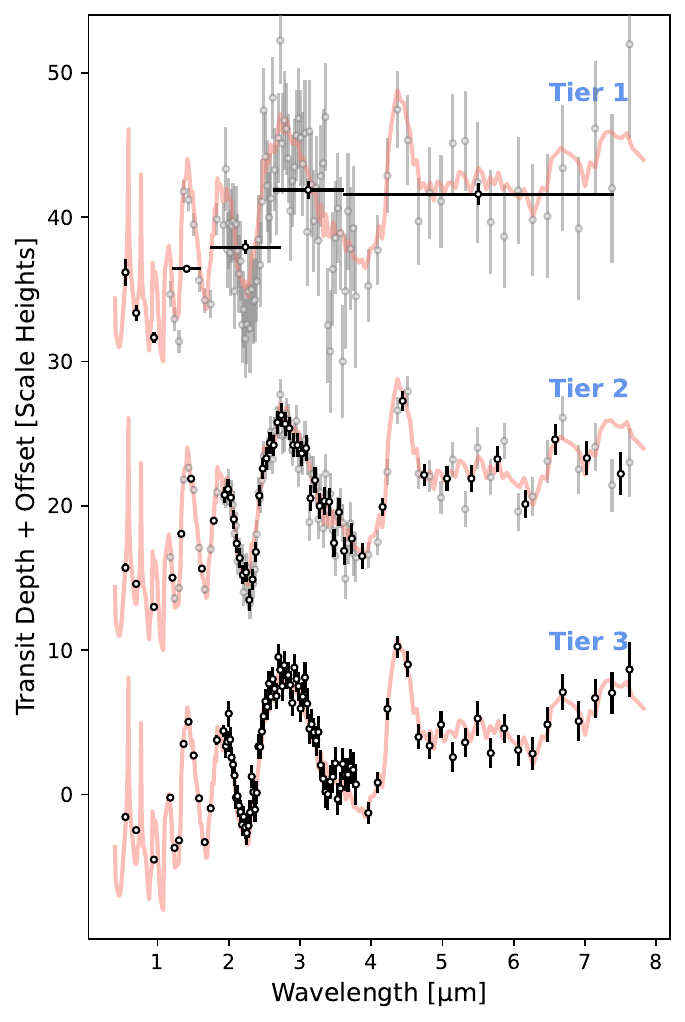}
    \caption{Similar to Figure~\ref{fig: Planet Spectra}, but showing observations of WASP-39\,b at different Ariel Tiers. Simulated data at the native resolution of each instrument are shown in grey faded points, and binned to the pre-defined Tier resolution in black (see text; \citealp{edwards_updated_2019}). The underlying atmosphere spectra are overplotted in red. Spectra are given an arbitrary vertical offset for ease of visualization.}
    \label{fig: Planet Spectra Tiers}
\end{figure}

We generate synthetic measurements in each wavelength bin, $\lambda_n'$, by sampling our atmospheric forward model and adding a noise offset sampled from a Gaussian with zero mean and standard deviation $\sigma_\delta(\lambda_n')$. We consider seven different observational S/N for each target, and 10 noise realizations at each precision. Figure~\ref{fig: Planet Spectra} depicts examples of simulated Tier 2 observations for random noise realizations of each of our three targets along with the underlying atmosphere forward models. Figure~\ref{fig: Planet Spectra Tiers} shows simulated observations of WASP-39\,b at different Tiers.

 
\section{Atmosphere Retrieval Analyses} 
\label{sec: Retrievals}

To assess the information content of our simulated spectra, we employ ``free retrievals''. Free retrievals parameterize most aspects of the atmospheric physics (e.g., temperature-pressure structure) and allow the abundances of each chemical species to vary independently \citep[see e.g.,][]{madhusudhan_atmospheric_2018}. We note that although we do not use the technical definition of the ``information content'' as defined in \citet{line_information_2012}, we effectively quantify the same thing --- the degree to which an observation increases our understanding of the atmosphere relative to (uninformative) priors.

We use \texttt{POSEIDON} for the retrievals, adopting an atmosphere setup (i.e., number of pressure layers, reference pressure, etc.) identical to that described in Section~\ref{sec: Forward Models} for the forward models. We assume the atmosphere to be isothermal and well-mixed, i.e., that the abundances of chemical species are constant with altitude. We reemphasize here that all retrievals are performed on the native resolution spectra.

We include the same atmospheric constituents for each planet as were listed in Section~\ref{sec: Forward Models} and Table~\ref{tab: Injected Abundances}. In each model we also include H$_2$-H$_2$ and H$_2$-He CIA, and allow for the possibility of aerosols using a standard parameterized ``cloud-haze'' prescription. This consists of an opaque, grey cloud deck placed at pressure $P_{\rm cloud}$, and the modified Rayleigh scattering slope as described in Section~\ref{sec: Forward Models}. Despite our injected cloud deck being spatially uniform (Section~\ref{sec: Forward Models}), we allow for the possibility of ``patchy'' clouds in the retrieval via the prescription of \citet{line_influence_2016}.

In each case, we fix the planet mass to the values in Table~\ref{tab: Planet Parameters} and fit for the scaled planetary radius. We assume an isothermal atmosphere temperature as transmission observations are not highly sensitive to the deep atmosphere thermal structure. However, to ensure that our results are not biased by our choice of temperature structure, we perform several tests using the \citet{madhusudhan_temperature_2009} parameterization and find nearly identical results to our isothermal case. We also test cases using the \citet{madhusudhan_temperature_2009} parameterization and a patchy cloud fraction of 0.4 for the data simulation, but again find virtually no difference in the results. Our final setup thus has a total of 15, 14, and 12 free parameters for WASP-39\,b, HAT-P-11\,b, and K2-18\,b, respectively. We sample the parameter space with \texttt{MultiNest} \citep{feroz_multinest_2009} using 1000 live points. For all three planets we retrieve on every noise realization at each observational S/N: 10 realizations per S/N $\times$ (7 + 7 + 7 S/N values) $\times$ 2 cloud models = 420 total retrievals. The prior ranges for each parameter are listed in Table~\ref{tab: Retrieval Priors}. 

\begin{figure*}
    \centering
    \includegraphics[width=0.8\linewidth]{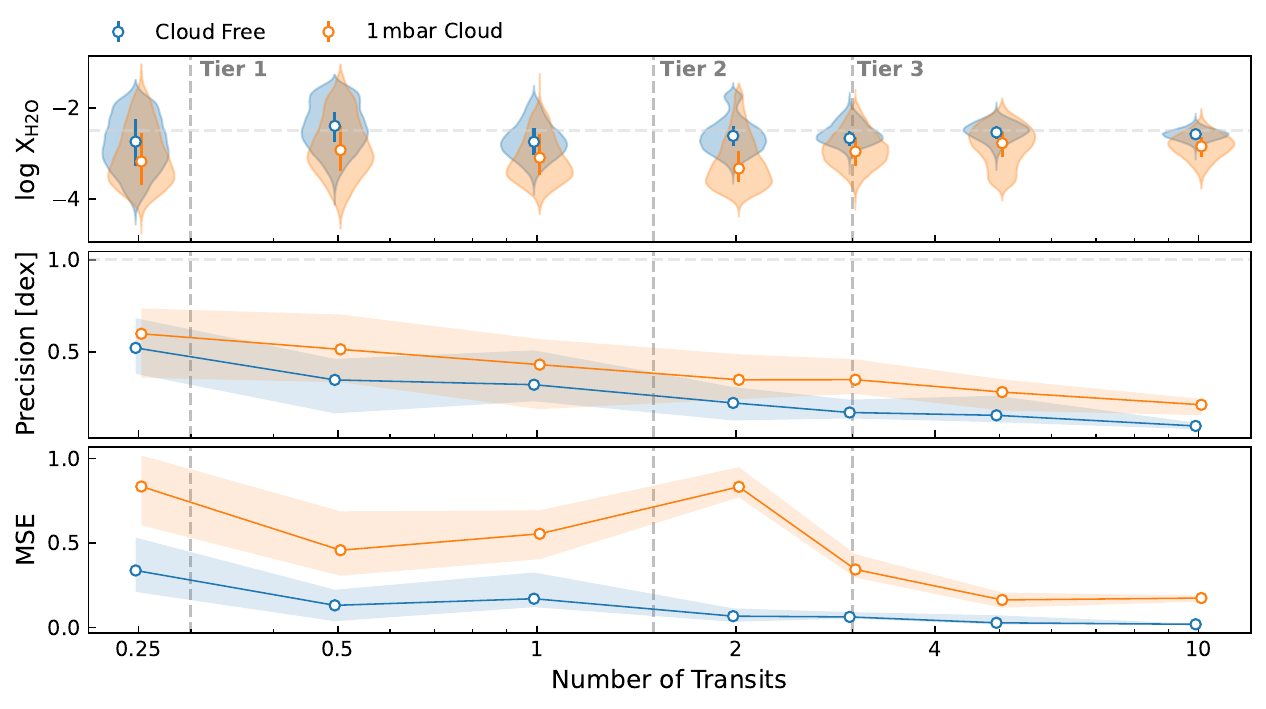}
    \caption{Constraints on H$_2$O in the atmosphere of a WASP-39\,b-like planet with Ariel as a function of observational S/N. 
    \emph{Top}: Retrieved abundance of H$_2$O as a function of observational S/N (i.e., number of stacked transits ---  assuming errors scale as 1/$\sqrt N$) for cloudy (orange) and cloud-free (blue) atmospheres. The injected abundance is denoted with a horizontal dashed grey line, and the number of transits corresponding to Tiers 1, 2, and 3, \citep[from][]{edwards_ariel_2022} with vertical dashed lines. The violin plots show the marginal posterior combined over all 10 realizations for each number of transits, and error bars denote the 1$\sigma$ credible range. Note that fractional transits are considered here to keep the results fully generalizable to other targets (see text).
    \emph{Middle}: Average precision on retrieved VMR as a function of observational S/N. The shaded envelope denotes the ranges of retrieved precisions across the 10 realizations. 
    \emph{Bottom}: Same as middle, but for the mean-squared error (MSE; Equ.~\ref{equ: MSE}). The MSE takes into account both the retrieved precision and any bias in the retrieved abundances.}
    \label{fig: W39b H2O}
\end{figure*}

In this study, our primary concern is ascertaining which chemical species can be confidently constrained, and with what precision, as a function of observational S/N (i.e., Ariel Tier). In the field, the ``detection'' of an atmospheric species is typically validated via Bayesian model comparison \citep[e.g.,][]{benneke_how_2013, thorngren_bayesian_2026}; one compares the Bayesian evidence value of a retrieval with a given species to one without (i.e., a leave-one-species-out test). Since this procedure would require an infeasible number of retrievals, we instead simply test whether the marginal posterior distribution for a given species is sufficiently bounded. To this end, we compare the ratio of the probability value at the peak of the marginal posterior to the value at the lower edge of the prior. For a Gaussian distribution, the ratio between the value at the distribution mean and the 99.7$\rm ^{th}$ percentile (i.e., in the 3-$\sigma$ tail) is $\gtrsim$0.011. If the marginal probability ratio is greater than this value for at least nine out of ten noise realizations, we consider a species to be well constrained at that observational S/N. 

We do note though several shortcomings with this method. The lower bound of the abundance priors are arbitrarily chosen \citep[e.g.,][]{thorngren_bayesian_2026}, and though our choices conform with the broader exoplanet literature, a different lower bound could change the number of species considered to be well-constrained. Moreover, we note that this method is also not robust to cases where a posterior is multi-modal --- which our method would consider to be constrained as long as one mode does not reach the lower prior bound.

Our findings are summarized in a series of plots showing the retrieved abundances, as well as quantifying the precision and bias (i.e., the difference between the injected and retrieved value) as a function of observational S/N for each planet and molecule. Figure~\ref{fig: W39b H2O} shows an example for H$_2$O in WASP-39\,b. Additional plots for select chemical species are included in Appendix~\ref{app: Additional Plots}, and plots for all other species are included in the Zenodo repository associated with this work\footnote{\url{https://zenodo.org/records/21456409}}. 

We use observational S/N instead of Ariel Tier as the independent variable here to explore setups that fall between the standard Tiers. For example, Tier 1 S/N for a given planet may be reached in one transit, and Tier 2 in three transits. But perhaps two transits are sufficient to constrain a chemical species of interest, and we wish to allow for this possibility. For each planet, we source the number of transits needed to reach each tier from \citet{edwards_ariel_2022}\footnote{\url{https://github.com/arielmission-space/Mission_Candidate_Sample}}. 

We note that for WASP-39\,b a single transit actually provides ``Tier-1.5" S/N, with Tier 1 S/N mathematically achievable with less than a single transit. Obviously, Ariel will only observe integer numbers of transits, however, we keep the fractional-transit definitions of the Tiers to keep the results generalizable.

\section{Results \& Discussion} 
\label{sec: Results}

As expected, the precision of retrieved abundances improves near-monotonically with observational S/N \citep[e.g.,][]{line_information_2012}. Moreover, we find that the retrieved precisions for cloudy atmospheres are generally lower than for cloud-free ones at a fixed S/N. This result makes intuitive sense as a cloud deck truncates the size of an atmospheric feature when observed in transmission, thereby decreasing the S/N of the feature itself compared to the cloud-free atmosphere (\citealp{fortney_effect_2005}; Figure~\ref{fig: Planet Spectra}).

Irrespective of the cloudiness of the atmosphere, our retrievals generally obtain the input abundance for each chemical species to within 1$\sigma$, though there are individual noise realizations for which the retrieved values deviate by $\sim$2$\sigma$. This is a similar finding to results on simulated JWST spectra \citep[e.g.,][]{welbanks_degeneracies_2019, davey_effect_2024} which found that there is still sufficient information content, even in cloudy spectra, to correctly retrieve atmospheric properties. As with most high-dimensionality atmospheric retrievals, there are some correlations between parameters that have competing effects on transmission spectra (e.g., Figure~\ref{fig: Example Corner}), and which can, in principle, result in biased atmospheric inferences \citep[e.g.,][]{benneke_atmospheric_2012, welbanks_degeneracies_2019}. However, we again find that our simulated Ariel observations generally have sufficient precision to break these degeneracies.

At the aggregate level, there is minimal systematic bias between the input and retrieved parameters for any species at any precision. Following \citet{rotman_enabling_2025} we combine both the bias and precision into a mean-squared error (MSE) value: 
\begin{equation}
    \mathrm{MSE} = b^2 + \sigma^2,
    \label{equ: MSE}
\end{equation}
where $\sigma^2$ is the posterior variance (i.e., the retrieved precision), and $b$ is the bias. In this way, if a retrieval settles on an incorrect VMR for a given species but is very confident about its solution, the retrieved precision will be high but the resulting MSE will be large due to the bias. However, since in our aggregate sample the bias is minimal, the retrieval precision drives the MSE to also decrease near-monotonically with S/N.

\begin{figure}
    \centering
    \includegraphics[width=\linewidth]{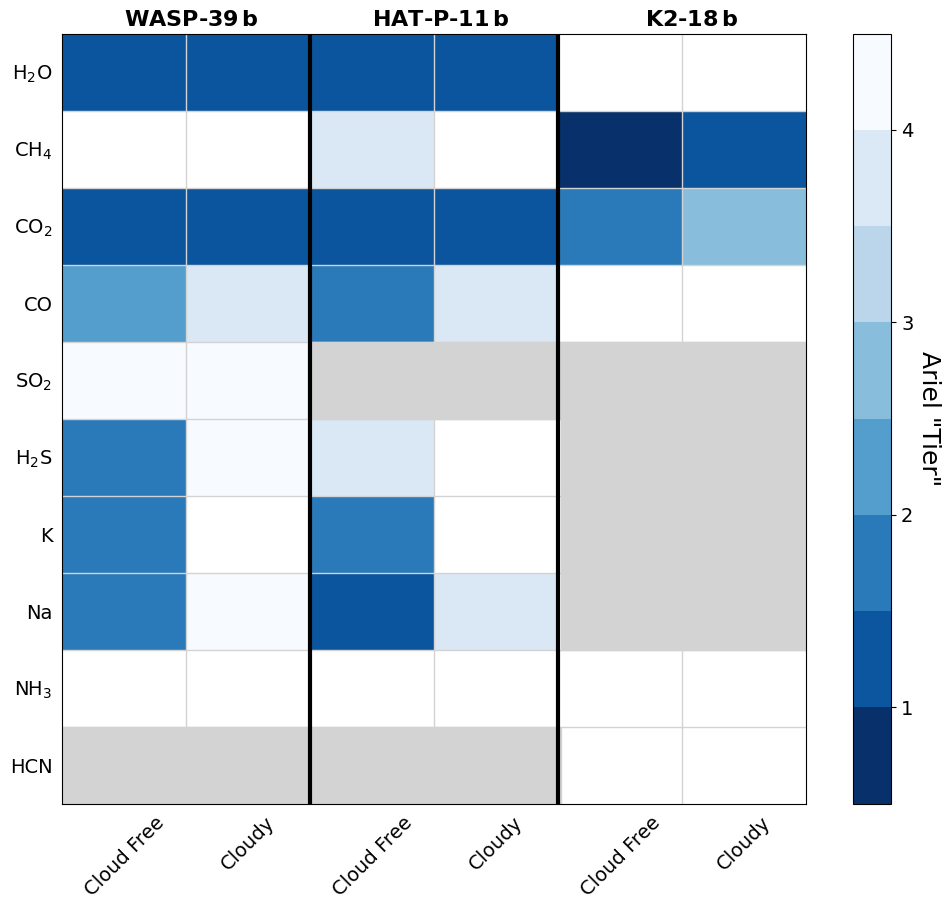}
    \caption{Summary of Ariel's constraining power across all tests conducted in this study. The colour denotes the Tier at which a given chemical species is consistently constrained for a given atmosphere setup. White shading means the species is not constrainable whereas grey means that it was not included in the atmosphere model. H$_2$O and CO$_2$ are well-constrained in a WASP-39\,b-like or HAT-P-11\,b-like planet at Tier 1 in both cloudy and cloud-free atmospheres, as are several other species at Tier 2 for cloud-free tests. CH$_4$ and CO$_2$ are the only molecules consistently identified for a K2-18\,b-like planet, with CH$_4$ being well-constrained at Tier 1. }
    \label{fig: Detection Limits Summary}
\end{figure}

In Figure~\ref{fig: Detection Limits Summary} we summarize the observational S/N necessary to obtain firm constraints, defined by the criteria in Section~\ref{sec: Retrievals}, for each chemical species and the three planets considered. 

For the WASP-39\,b-like planet, we find that firm constraints with $<$1\,dex precision are obtainable with Tier 1-quality observations for some of the main atmospheric constituents like H$_2$O and CO$_2$ in cloud-free atmospheres. Cloudy atmospheres also allow for H$_2$O and CO$_2$ constraints at Tier 1, though with a slightly lower precision. H$_2$O and CO$_2$ are two of the dominant carriers of O and C in hot exoplanet atmospheres, and are thus major necessary pieces for constraining fundamental properties like metallicity and C/O ratio \citep[e.g.,][]{moses_disequilibrium_2011, madhusudhan_co_2012}. This means that for a typical hot-Saturn or hot-Jupiter, Tier 1-quality observations could already give important insights into atmospheric chemistry and begin to uncover population-level trends in, e.g., metallicity. 

A caveat to this is that CO, which should be the dominant carrier of C in a WASP-39\,b-like planet, remains unconstrainable until ``Tier 2.5"-quality observations if the atmosphere is cloud-free and Tier 3 if cloudy. This is largely due to CO and CO$_2$ having overlapping opacity (e.g., Figure~\ref{fig: Spectrum Contributions}), making the weaker CO feature more difficult to identify in the presence of a strong CO$_2$ band. This is also something that plagues JWST observations, with CO most often being directly detected in low-resolution spectra in lower-metallicity atmospheres without significant CO$_2$ \citep[e.g.,][]{meech_bowie-align_2025, kirk_bowie-align_2025, changeat_synergetic_2025, claringbold_bowie-align_2026}. 

We, therefore, run an additional test artificially decreasing the abundance of CO$_2$ to log~VMR$\sim$$-$6 such that it no longer features strongly in the spectrum. Our retrieval analyses now show that both CO and CO$_2$ can be jointly constrained with Tier 1.5-quality observations in a cloud-free atmosphere. However, Tier 2 precision is still required if the atmosphere is cloudy. 

Irrespective of the CO$_2$ feature strength, minor species like H$_2$S can be constrained with Tier 1.5--2-quality observations, and the alkalis Na and K at Tier 1.5 in cloud-free atmospheres, but will be difficult to detect, even at Tier 3 if the atmosphere is cloudy. SO$_2$ remains undetected in both cases, even at Tier 3. This is largely because, even at the native resolution of AIRS Ch1, the SO$_2$ feature at $\sim$4\,µm is covered by only two wavelength bins, making it highly-sensitive to the particular noise realization.

To summarize, for a WASP-39\,b-like planet, Tier 1-quality observations are already sufficient to provide firm constraints on H$_2$O and CO$_2$ via free retrievals in cloudy or cloud-free atmospheres and begin to construct population-level trends. There is a minimal increase in precision moving from Tier 1 to Tier 2 in these major species, although other important molecules like CO become identifiable in cloud-free atmospheres at ``Tier 2.5". Constraining minor species like H$_2$S and alkalis generally requires at minimum Tier 1.5--2 observations.

For a warm-Neptune like HAT-P-11\,b, the results are qualitatively similar --- a benefit and outcome of defining Tiers based on observational S/N. H$_2$O and CO$_2$ are again well constrained at Tier 1 with $<$1\,dex precision irrespective of cloud cover. CO can also be constrained at ``Tier 1.5'' if cloud-free, likely due to its significantly higher abundance here compared to the hot-Saturn. However, unlike for the hot-Saturn, identifying species beyond these three generally requires much higher-tier data; Tier 3+ is necessary to constrain H$_2$S or CH$_4$ if the atmosphere is cloud-free, and they remain unconstrained if cloudy. When robustly identified, though, abundances are well-constrained with precisions of $\sim$1\,dex in most cases. 

The case for K2-18\,b is weaker, with CH$_4$ and CO$_2$ being the only constrainable molecules at any Tier. CH$_4$ should be identifiable in Tier 1 for a cloud-free atmosphere, whereas Tier 2 is necessary if cloudy. CO$_2$ is well constrained with a ``Tier 1.5''-quality dataset if cloud-free, but remains undetected at any tier if cloudy. This highlights the challenge of studying the atmospheres of smaller and colder planets with Ariel. 

Another consideration on this front is the number of transits required to reach a given S/N. Whereas for a WASP-39\,b-like planet, Tier 2 S/N is obtained in two transits, 14 transits are necessary for the equivalent S/N in HAT-P-11\,b due to the smaller size of its expected atmospheric features. This is not beyond the realm of possibility given Ariel's four-year primary mission, though certainly on the upper-edge of what might be feasible in the Tier 2 survey. However, similar planets with either brighter host stars or larger expected atmosphere features could be achievable in fewer transits and therefore be an important addition to the Tier 2 survey. For K2-18\,b, though, nearly 100 transits are necessary to reach Tier 2 S/N. This means that unless planets similar to K2-18\,b are discovered around significantly brighter stars, the chemistry of temperate sub-Neptunes will be a challenge to constrain with Ariel. 

Although, as mentioned above, we wanted to avoid running Bayesian model comparison analyses to quantify detection significances due to the computational cost of such an endeavour, to provide extra credence to our results we run several leave-one-species-out spot checks on individual planets and noise realizations. In general, species which are well constrained by our method are also detected via the Bayesian leave-one-out test with ``decisive'' evidence (odd ratio $>$ 350:1, $\Delta \ln~B>5.91$). The one exception is Na which in several cases is well constrained via our test, but does not reach the ``decisive'' evidence threshold in the Bayesian leave-one-out analysis. This is perhaps unsurprising as the narrow Na line is probed solely by a singular point of VISPhot photometry. Na (and K for that matter), though, is only a minor species and not a strong tracer of first order atmospheric properties like metallicity and C/O ratio which Ariel aims to measure.

\section{Conclusions} 
\label{sec: Conclusions}

In this work, we explored how well various key chemical species can be constrained in Ariel transmission spectra as a function of Tier (i.e., observational S/N). Via a suite of free retrievals, we ascertained which Tier is necessary for firm constraints in three benchmark planets. We summarize our findings for each planet below.

\begin{itemize}
    \item For a WASP-39\,b-like hot-Saturn, Tier 1-quality observations are sufficient for $\lesssim$1\,dex constraints on H$_2$O and CO$_2$ in cloudy or cloud-free atmospheres. There is an incremental increase in precision at Tier 2. CO, as well as secondary species like H$_2$S, Na, and K are identifiable in cloud-free atmospheres with observations at Tier 2+ quality.
    
    \item For a HAT-P-11\,b-like warm-Neptune, H$_2$O, and CO$_2$ are again constrainable at Tier 1 irrespective of the presence of grey clouds. CO can also be constrained at Tier 1 for cloud-free atmospheres, but requires Tier 3+ if cloudy. Secondary species generally require Tier 3 if cloud-free or remain unconstrained if cloudy. The number of transits necessary to reach this precision might be a challenge to fit into Ariel's Tier 2 survey.

    \item For a K2-18\,b-like temperate sub-Neptune, only CH$_4$ and CO$_2$ are potentially identifiable, even at Tier 3. CH$_4$ can be constrained at Tier 1 in cloudy and cloud-free atmospheres. CO$_2$ requires Tier 1.5, or 2.5+ if cloudy. The number of transits necessary to reach this precision is likely prohibitive. Comparable targets with larger atmospheric features and/or brighter host stars may still remain within Ariel's reach \citep[e.g.,][]{changeat_synergetic_2025}.
    
\end{itemize}

In general, for the giant planets which are expected to make up the bulk of the MRS, we find that Tier 1 already provides a solid foundation to begin to identify population-level trends in exoplanet atmospheric chemistry. There is moderate information gain in terms of constraints on already-detected species when moving to Tiers 2 and 3. The real benefits come from unlocking a wider range of species, allowing for more precise constraints on bulk properties like metallicity and C/O. In this light, Tier 3 observations should be prioritized wherever possible for giant planets over Tier 2 in order to unlock the contributions of key species like CO in cloudy atmospheres.  

Our results demonstrate that Ariel's Tier 1 survey need not be a ``vetting sample'' to determine planets suitable for the Tier 2 survey, but could themselves yield important constraints on atmospheric chemistry, in line with the simulations of \citet{mugnai_alfnoor_2021}. Insights can be gained into population-level trends even when considering only a limited number of chemical species \citep[e.g.,][]{welbanks_massmetallicity_2019}. In this light, Ariel's Tier 1, survey, with a predicted sample of $\sim$1000 planets, has the potential to turbo charge these initial insights and perform comparative exoplanetology on a scale never seen before.

\begin{acknowledgments}
M.R.\ acknowledges funding from the Natural Sciences and Engineering Research Council of Canada (NSERC), as well as the Canadian Space Agency (CSA).
This research has made use of the NASA Exoplanet Archive, which is operated by the California Institute of Technology, under contract with the National Aeronautics and Space Administration under the Exoplanet Exploration Program. 
\end{acknowledgments}

\vspace{5mm}
\facilities{Exoplanet Archive \citep{christiansen_nasa_2025}}

\software{\texttt{astropy} \citep{astropy:2013, astropy:2018}, 
\texttt{ipython} \citep{PER-GRA:2007},
\texttt{matplotlib} \citep{Hunter:2007},
\texttt{numpy} \citep{harris2020array},
\texttt{POSEIDON} \citep{macdonald_hd_2017, macdonald_poseidon_2023},
\texttt{pymultinest} \citep{buchner_statistical_2016},
\texttt{scipy} \citep{2020SciPy-NMeth}
}

\appendix

\section{Additional Plots \& Tables}
\label{app: Additional Plots}

\begin{figure*}
    \centering
    \includegraphics[width=0.75\linewidth]{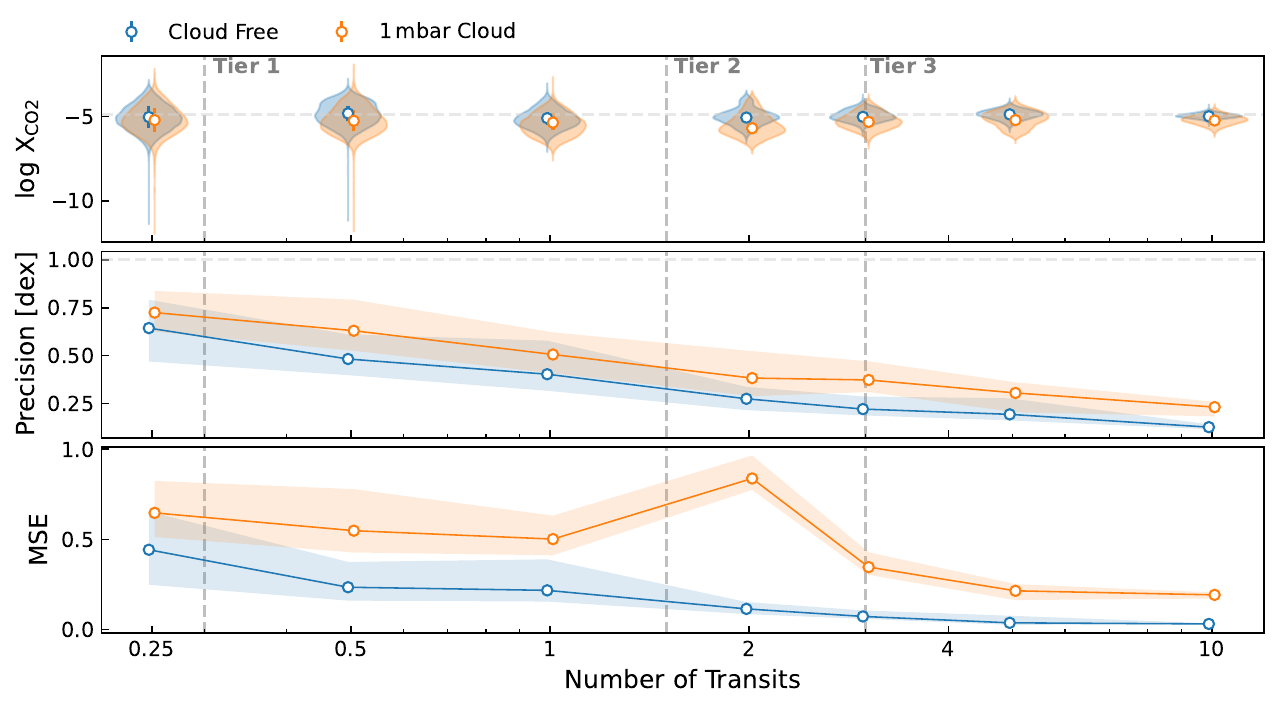}
    \caption{Same as Figure~\ref{fig: W39b H2O}, but showing trends for CO$_2$ in WASP-39\,b. In the lower-S/N cases, the violins have tails extending to the lower edge of the VMR prior reflecting one particular noise realization which was unable to constrain CO$_2$ according to our definition in Section~\ref{sec: Retrievals}. However, since nine other realizations could constrain CO$_2$, we still consider it to be well constrained at these observational S/N values.}
    \label{fig: W39b CO2}
\end{figure*}

\begin{figure*}
    \centering
    \includegraphics[width=0.75\linewidth]{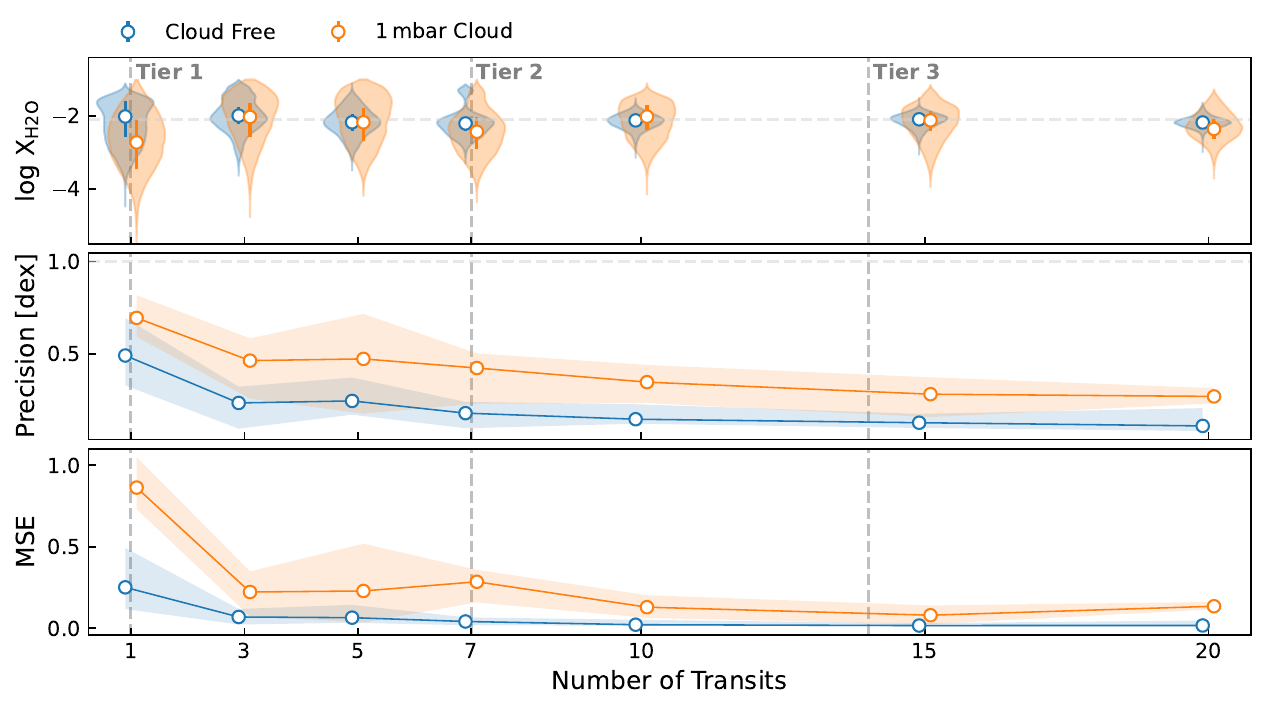}
    \caption{Same as Figure~\ref{fig: W39b H2O}, but showing trends for H$_2$O in HAT-P-11\,b.}
    \label{fig: HP11b H2O}
\end{figure*}

\begin{figure*}
    \centering
    \includegraphics[width=0.75\linewidth]{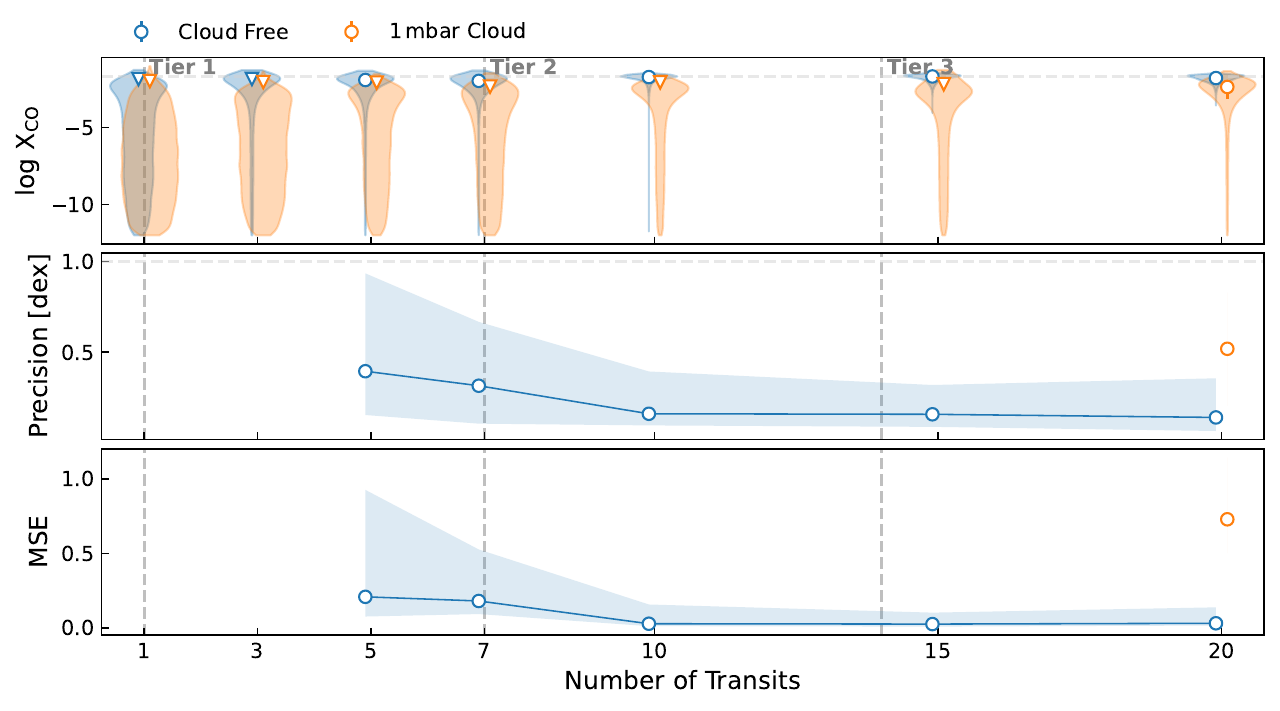}
    \caption{Same as Figure~\ref{fig: W39b H2O}, but showing trends for CO in HAT-P-11\,b. In the top panel, triangles denote a 3$\sigma$ upper limit in cases where a species is not constrained in at least 90\% of noise realizations. In the lower two panels, no precision or MSE is shown in such cases. }
    \label{fig: HP11b CO}
\end{figure*}

\begin{figure*}
    \centering
    \includegraphics[width=0.75\linewidth]{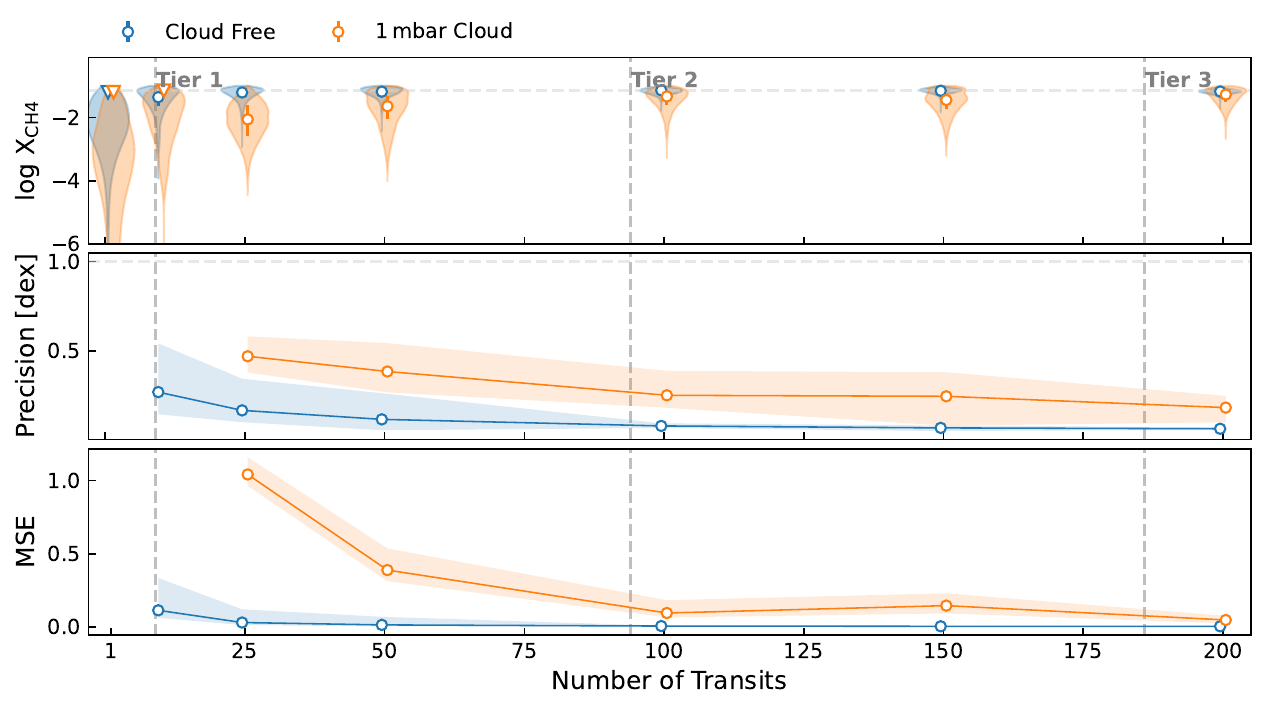}
    \caption{Same as Figure~\ref{fig: HP11b CO}, but showing trends for CH$_4$ in K2-18\,b.}
    \label{fig: K218b CH4}
\end{figure*}

\begin{figure*}
    \centering
    \includegraphics[width=0.9\linewidth]{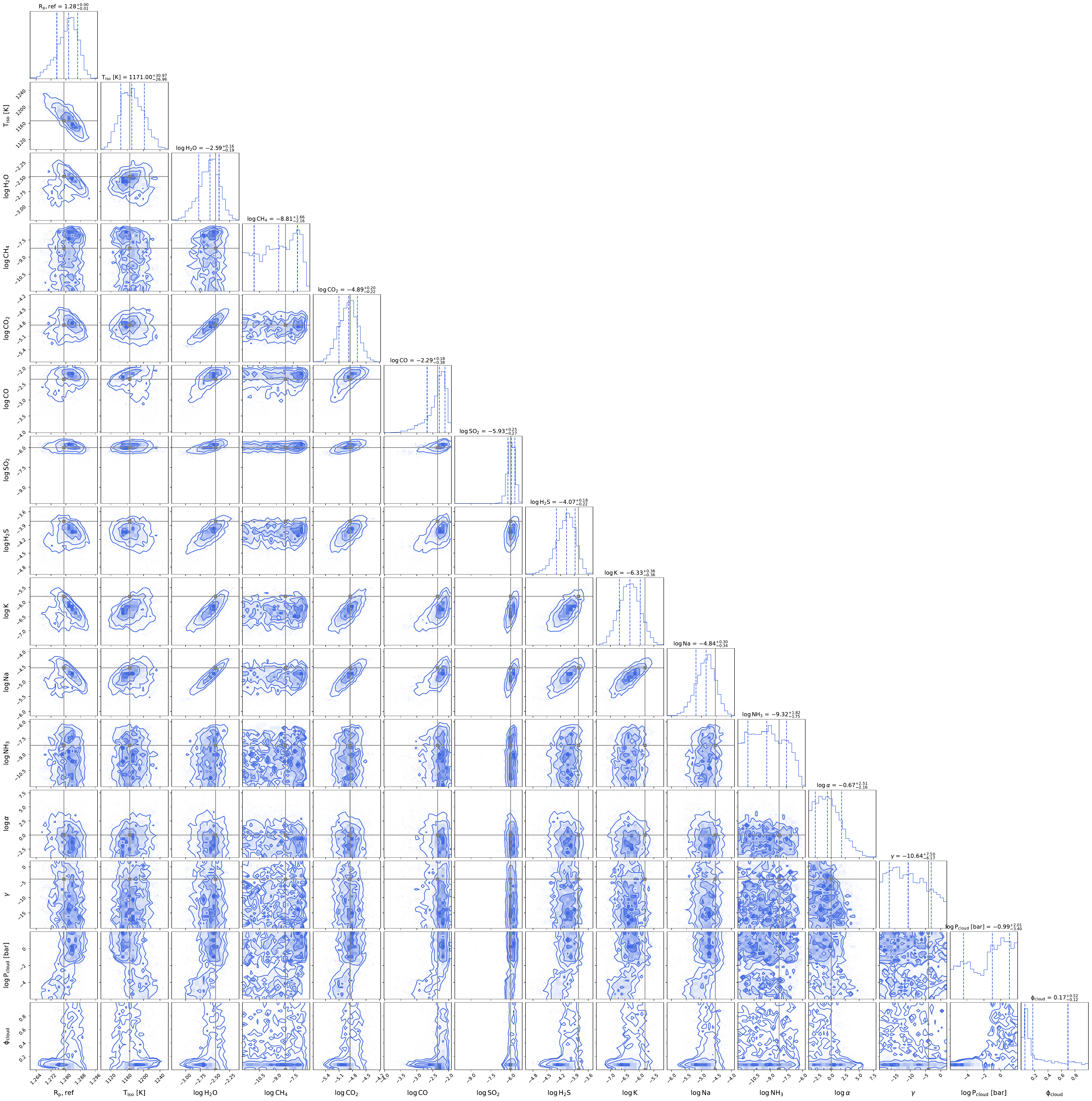}
    \caption{An example corner plot for one noise realization of the cloud-free, five-transit WASP-39\,b simulations. Solid grey lines in the marginal distributions represent the injected values. Dashed blue lines show the posterior median and 1-$\sigma$ range.}
    \label{fig: Example Corner}
\end{figure*}

Here, we show plots analogous to Figure~\ref{fig: W39b H2O} for a selection of other major species in the three benchmark planets. Figure~\ref{fig: W39b CO2} shows trends in CO$_2$ in WASP-39\,b, Figures~\ref{fig: HP11b H2O} and \ref{fig: HP11b CO} show H$_2$O and CO for HAT-P-11\,b, and Figure~\ref{fig: K218b CH4} shows CH$_4$ for K2-18\,b. Results for all other molecules considered are included in the associated Zenodo archive\footnote{\url{https://zenodo.org/records/21456409}}. Figure~\ref{fig: Example Corner} also shows an example corner plot for one noise realization of the cloud-free WASP-39\,b simulations.

Table~\ref{tab: Injected Abundances} summarizes the atmospheric abundances injected into forward models of each of the three planets considered in this study, and Table~\ref{tab: Retrieval Priors} shows the retrieval priors. 

\begin{deluxetable}{c|ccc}
 \centering
 \tabletypesize{\small}
 \label{tab: Injected Abundances}
 \tablecaption{Injected Chemical Abundances}
 \tablehead{Species & WASP-39\,b & HAT-P-11\,b & K2-18\,b}
    \startdata
     H$_2$O & $-$2.49 & $-$2.10 & $-$6.87 \\
     CO$_2$ & $-$4.85 & $-$3.27 & $-$2.05 \\
     CO & $-$2.35 & $-$1.68 & $-$5.00 \\
     CH$_4$ & $-$8.21 & $-$6.31 & $-$1.89 \\
     Na & $-$4.54 & $-$3.91 & - \\
     K & $-$5.80 & $-$5.50 & - \\
     NH$_3$ & $-$8.17 & $-$7.05 & $-$5.00\\
     SO$_2$ & $-$6.00 & - & - \\
     H$_2$S & $-$3.81 & $-$4.35 & - \\
     HCN & - & - & $-$5.00\\
    \enddata
    \tablecomments{Abundances are log~VMR and assumed to be vertically uniform throughout the terminator atmosphere.
    The injected abundances follow from chemical equilibrium, as described in Section~\ref{sec: Forward Models}.
    A - symbol indicates that the species was not included in a given model.}
\end{deluxetable}

\begin{deluxetable}{c|c}
 \centering
 \tabletypesize{\small}
 \label{tab: Retrieval Priors}
 \tablecaption{Retrieval Priors}
 \tablehead{Parameter & Prior Range}
    \startdata
     log VMR & $\mathcal{U}$[$-$12, $-$1] \\
     $\rm \log P_{cloud}$ [bar] & $\mathcal{U}$[$-6$, 2] \\
     $\rm \log \alpha$ & $\mathcal{U}$[$-$4, 8] \\
     $\rm \gamma$ & $\mathcal{U}$[$-20$, 2] \\
     $\rm \phi_{cloud}$ & $\mathcal{U}$[0, 1] \\
     $\times$R$\rm _{p}$ & $\mathcal{U}$[0.75$\times$R$\rm _p$, 1.25$\times$R$\rm _p$] \\ 
     T$\rm _{iso}$ [K] & $\mathcal{U}$[100, 2000] \\ 
    \enddata
    \tablecomments{$\mathcal{U}$ denotes a uniform prior on the specified range.
    VMR prior ranges for all chemical species are the same.}
\end{deluxetable}

\bibliography{main}{}
\bibliographystyle{aasjournal}

\end{document}